\documentclass[12pt,3p,sort&compress]{elsarticle}

\usepackage{multicol}
\usepackage{color}
\usepackage{xspace}
\usepackage{graphicx} 
\usepackage{multirow}
\usepackage{array} 
\usepackage{amsmath}
\usepackage{amssymb}

\def\bs{\boldsymbol}

\def\yb{\bar{\boldsymbol y}}
\def\zb{\bar{\boldsymbol z}}

\def\k{{\boldsymbol k}}

\def\x{{\boldsymbol x}}
\def\y{{\boldsymbol y}}
\def\B{{\boldsymbol B}}

\def\r{{\boldsymbol r}}
\def\z{{\boldsymbol z}}

\def\G{{\cal G}}
\def\U{\mathcal{U}}

\def\P{\mathcal{P}}
\def\bkappa{{\boldsymbol \kappa}}
\def\bbkappa{\bar{\boldsymbol \kappa}}

\newcommand{\beq}{\begin{eqnarray}}
\newcommand{\eeq}{\end{eqnarray}}
\newcommand{\be}{\begin{eqnarray*}}
\newcommand{\ee}{\end{eqnarray*}}

\begin{document}

\begin{frontmatter}

\title{Coherence effects between the initial and final state radiation in a dense QCD medium}

\author[USC,OSU]{Mauricio Martinez}

\address[USC]{Departamento de F\'isica de Part\'iculas and IGFAE, 
Universidade de Santiago de Compostela, 
E-15706 Santiago de Compostela, 
Galicia-Spain}
\address[OSU]{Department of Physics, The Ohio State University, Columbus, OH 43210, USA}

\begin{abstract}

In these proceedings contribution we review our recent results of medium modifications to the coherence pattern between the initial and final state radiation. We study single gluon production off a highly energetic parton that undergoes a hard scattering and subsequently crosses a dense QCD medium of finite size. Multiple scatterings lead to a partial suppression of the interferences between different emitters that opens the phase space for large angle emissions. We briefly discuss the generalization of this setup to describe gluon production in the case of pA collisions by including finite length and energy corrections. The configuration studied here may have phenomenological consequences in high-energy nuclear collisions.

\end{abstract}


\end{frontmatter}
\section{Introduction}
\label{sec: intro}

Color coherence effects in jet physics in the vacuum have been extensively studied at theoretical and experimental level in the last decades. Different observables such as jet shapes, intrajet activity and multiplicity have confirmed their relevance in high energy hadronic collisions~\cite{Braunschweig:1990yd,Abbiendi:2002mj,Abe:1994nj}.

Color coherence addresses the question about to what extent different emitters in the cascade act independently. To answer this question let us consider soft gluon radiation in a deep inelastic process for the color singlet configuration (t channel). In this case, the probability to emit a gluon from any of the emitters is \cite{Dokshitzer:1991wu}
\beq
\label{vacemiss}
\langle dN_{in}\rangle_{\phi}=\frac{\alpha_s C_F}{\pi} \frac{d\omega}{\omega} \frac{d(\cos\theta)}{1-\cos\theta}\Theta (\cos\theta-\cos\theta_{qq})\,,
\eeq
where $\theta$ is the angle of emitted gluon and $\theta_{qq}$ is the angle between the incoming and the outcoming parton due to hard scattering with the virtual photon. Eq.(\ref{vacemiss}) shows us that gluon radiation is going to be confined inside the cone defined by the opening angle $\theta_{qq}$ along the emitter. This simple example is the basis of the {\it angular ordering} prescription in jet evolution. A simple and intuitive probabilistic approach of the partonic cascade produced in hadronic collisions emerges from this simple exercise \cite{Dokshitzer:1991wu}. Perhaps the most relevant experimental observation which confirms this phenomena is the suppression of the soft sector in the jet fragmentation function (the hump-backed plateau)~\cite{Braunschweig:1990yd,Abbiendi:2002mj,Abe:1994nj}.

The study of medium modifications to the branching process has been addressed recently in a series of papers \cite{MehtarTani:2010ma,CasalderreySolana:2011rz,MehtarTani:2011gf,MehtarTani:2012cy} by considering the radiation pattern of a $q\bar{q}$ antenna immersed in a QCD medium. In these proceedings, we briefly review an extension of color coherence studies inside a QCD medium  to a space-like (t-channel) scattering process\cite{Armesto:2012qa,Armesto:2013fca}. We also mention how to generalize our studies to a more realistic scenario such as pA collisions.  

\section{Soft gluon radiation from Classical Yang-Mills Equations}
\label{sec:CYM}
In this section we provide a brief review of the calculation of soft gluon radiation by solving the Classical Yang Mills (CYM) Eqs. For a complete description of the technical details of this calculation, we refer to the interested reader to Ref.~\cite{Armesto:2013fca} and references therein. Within the semiclassical method the radiated gluon $A_\mu^a$ is treated as a fluctuation on top of a static background gauge field $A_\mathrm{med}$ (the medium). In the presence of the  current generated by the incoming and outcoming quark, $J^\mu$, the CYM equations read $\left[D_\mu, F^{\mu\nu} \right] = J^\mu$, where $D_\mu = \partial_\mu - i g A_\mu$, $F_{\mu\nu} = \partial_\mu A_\nu - \partial_\nu A_\mu - i g [A_\mu, A_\nu]$ and $g$ is the gluon coupling constant. For the case of interest, we have two currents associated to the incoming quark\footnote{For the outoming quark one must change the argument of the Heaviside step function $\Theta(-t)\to\Theta(t)$ and the 4-momentum $p^\mu\to \bar{p}^\mu$.} 
is given by $J^{\mu,a}_{inc} = g \, p^\mu /E \Theta(-t)\delta^{(3)}(\vec x-\vec p/E t) \Theta(-t)Q_{inc}^a$ , where $p^\mu=(E,\vec p)$ and $Q_{inc}^q$ the color charge. The color current follows the continuity equation $[D_\mu, J^\mu] = 0$ with $J = J_{inc} + J_{out}$. Finally, the amplitude of emitting a gluon with momentum $k \equiv (\omega, \vec k)$ is given by the recursion relation
\beq
\mathcal{M}_\lambda^a(\vec k) = \lim_{k^2\to 0} -k^2 A_\mu^a (k) \epsilon_\lambda^\mu(\vec k) \,,
\eeq
where $\epsilon_\lambda^\mu(\vec k)$ is the gluon polarization vector. Our calculation is performed in the light cone gauge where $A^+=0$. The gluon spectrum can be calculated by taking the square of the scattering amplitude. Nevertheless, it is necessary to indicate how to perform the medium averages. For simplicity we consider that the medium averages are performed within the harmonic oscillator approximation.

\begin{figure}[t]
\begin{center}
\includegraphics[width=12cm]{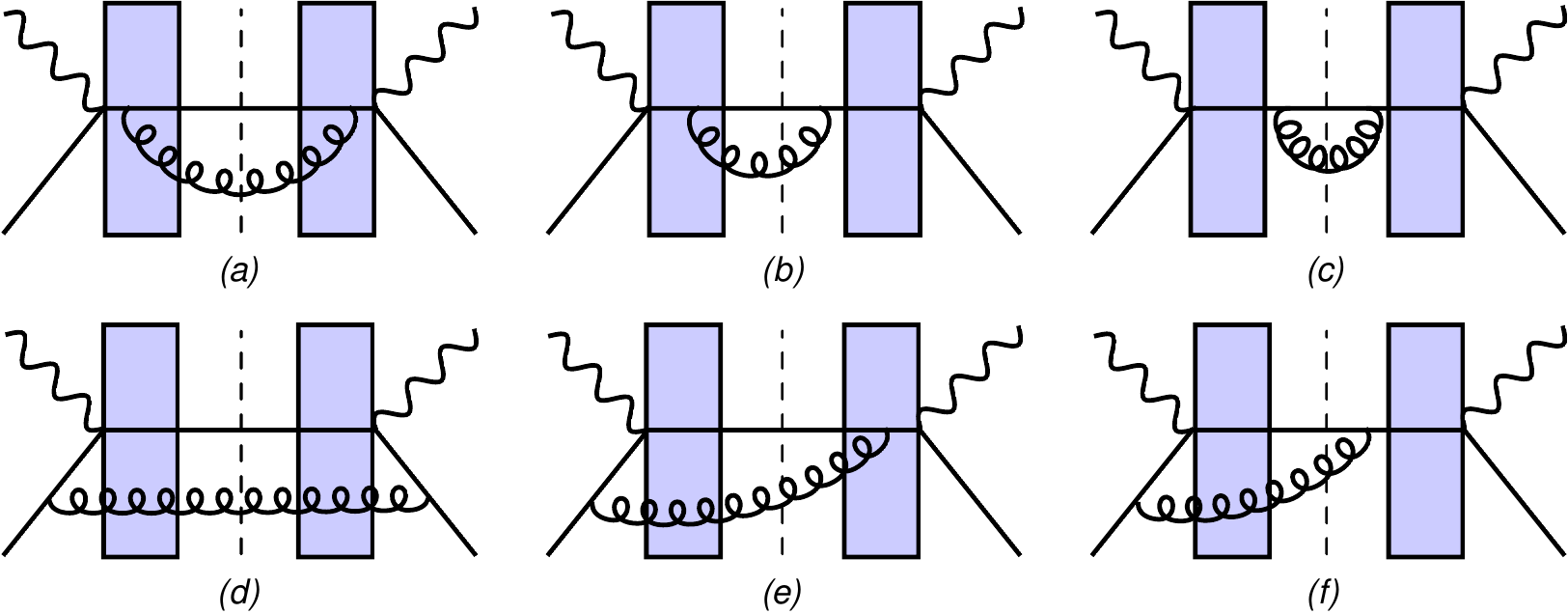}
\end{center}
\caption{Different components of the gluon spectrum:  (a) the ``\emph{in-in}" component, (b) the ``\emph{in-out}" component, (c) the ``\emph{out-out}" component, (d) the ``\emph{bef-bef}" component, (e) the ``\emph{bef-in}" component, and (f)  the ``\emph{bef-out}" component.}
\label{fig:spec}
\end{figure}
%

\subsection{The gluon spectrum}

We split the gluon spectrum in six different components according to the longitudinal position where the gluon is emitted in the amplitude and the complex conjugate with respect the hard scattering: the direct emissions of the incoming parton (``\emph{bef-bef}") and the outgoing one (``\emph{in-in}",``\emph{in-out}" and ``\emph{out-out}"), and the interferences between both emitters (``\emph{bef-in}" and ``\emph{bef-out}") (see Fig.\ref{fig:spec})~\cite{Armesto:2013fca}.  The gluon spectrum then reads
\beq
 k^+ \frac{dN_{tot}}{d^3k}= \sum_{i=1}^{6}  k^+ \frac{dN_i}{d^3k}\ . 
\eeq
In the following we briefly describe the properties of each component of the gluon spectrum.

\subsubsection{Direct emissions}
Here we describe some of the main properties of the direct emissions of the incoming (``\emph{bef-bef}") and outgoing parton (``in-in", ``in-out" and ``out-out")~\cite{Armesto:2013fca}. Some of these results are known in the literature.

\begin{itemize}
\item [{\bf (i)}] $\textbf{``\emph{Bef-bef}"}$:  this term can be written as
\beq
\label{eq:befcomp}
k^+ \frac{dN_{bef-bef}}{d^3k}
&=&\frac{\alpha_s\,C_F}{\pi^2}\, \text{Re}\Biggl\{\int \frac{d^2\k'}{(2\pi)^2}\frac{\P(\k'-\bkappa, L^+)}{\k'^2}\,
\Biggr\}\,,
\eeq
where $\bkappa$ is the transverse momentum of the gluon relative. The function $\P (\k,\xi)$ is defined as
\begin{equation*}
\label{eq:Pdef}
\P (\k, \xi)=\frac{4 \pi}{\hat q \xi}\exp\Biggl[-\frac{\k^2}{\hat q \xi}\Biggr]\,.
\end{equation*}
$\P(\k,\xi)$ is the probability of having accumulated a certain transverse momentum $\k^2$ while traversing a longitudinal distance $\xi$. The ``\emph{bef-bef}" contribution (\ref{eq:befcomp}) is understood as a two-step process: a gluon with momentum $\k'$ is emitted in the remote past by the incoming parton, the term $1/\k'^2$ in Eq. (\ref{eq:befcomp}), and afterwards the gluon follows a random walk while crossing the medium described by $\P(\k'-\bkappa, L^+)$. Gluons from the initial state get their momenta reshuffled and in average, they acquire the maximal amount of accumulated transverse momentum $Q_s^2=\hat q L^+$ where $\hat q$ is a medium property related with the measure of the transverse momenta per unit lenght. In addition, the multiplicity of gluons associated does not change its value, instead its angular distribution gets smeared out to large angles~\cite{Armesto:2013fca}. 

\item [{\bf (ii)}] $\textbf{``\emph{In-in}"}$: this contribution corresponds to the emission completely inside the medium by the outcoming parton. The analytical expression is calculated and give us
\beq
\label{eq:inmed}
\hspace{-1cm}k^+ \frac{dN_{in-in}}{d^3k}&=&  \frac{\alpha_s C_F}{2\pi^2}
\,\frac{1}{(k^+)^2}\,\text{Re}\Biggl\{
2 i k^+\,\int_0^{L^+}dy^+\int\frac{d^2\k'}{(2\pi)^2}\,\P(\k'-\bbkappa,L^+-y^+)\nonumber\\
&\times&\exp\Bigl((1-i)\frac{\k'^2}{2k_f^2}\tan\bigl(\Omega y^+\bigr)\Bigr)
\Biggr\},
\eeq
where $k_f^2=\sqrt{k^+ \hat q}$. Eq. (\ref{eq:inmed}) shows that the medium-induced component is also understood as a two-step process. First, the quantum emission of a gluon with momentum $\k'$  at certain time $\tau_f\sim |\Omega|^{-1}$ (the exponential term in the second line of Eq. (\ref{eq:inmed})) and afterwards, the subsequent brownian motion of the gluon along the remaining path through the medium which is described by $\P(\k'-\bbkappa,L^+-y^+)$. The emission spectrum peaks around $k_f$ which corresponds to the amount of  momentum accumulated during its formation time $\tau_f$. The medium-induced component (\ref{eq:inmed}) scales with the length of the medium $L^+$~\cite{Armesto:2013fca}. 

\item [{\bf (iii)}] $\textbf{``\emph{In-out}"}$: this term takes into account the medium-vacuum interference of the final state radiation and it reads as
\beq
\label{eq:inout-exact}
k^+ \frac{dN_{in-out}}{d^3k}=-2\frac{\alpha_s\,C_F}{\pi^2}\frac{1}{\bbkappa^2}\text{Re}\Biggl\{
1-\exp\Biggl[-i\frac{\bbkappa^2}{2k^+\Omega}\tan (\Omega L^+)
\Biggr]
\Biggr\}.
\eeq
This contribution becomes important when emissions take place near to the boundary of the medium~\cite{Armesto:2013fca}.

\item [{\bf (iv)}] $\textbf{``\emph{Out-out}"}$: this contribution takes place completely outside the medium due to bremmstrahlung of the outcoming parton and is given by~\cite{Armesto:2013fca}
\beq
\label{eq:outout-2}
k^+ \frac{dN_{out-out}}{d^3k}=\frac{\alpha_s\,C_F}{\pi^2}\,\frac{1}{\bbkappa^2}\,.
\eeq
\end{itemize}

\subsubsection{Interferences}
\label{subsec:interf}
The information about the colour correlation between both emitters is carried out by the interference terms ``\emph{bef-in}" and ``\emph{bef-out}".  The color correlation between both emitters will depend on whether the QCD medium is able to resolve the quark-gluon system or not. This is analogous to the studied case of the $q\bar q$ antenna immersed in a QCD medium~\cite{MehtarTani:2010ma,CasalderreySolana:2011rz,MehtarTani:2011gf,MehtarTani:2012cy,MehtarTani:2011tz}, with $r_\perp$  the transverse $q\bar q$ dipole size. However, the transverse size of the quark-gluon system depends on the accumulated transverse momentum of the gluon at a certain longitudinal position inside the medium due to its brownian motion~\cite{Armesto:2013fca}. 

\begin{itemize}
\item  [{\bf (i)}] $\textbf{``\emph{Bef-in}"}$: this term can be expressed as follows
\beq
\label{eq:befin-comp}\hspace{-1cm}
k^+ \frac{dN_{bef-in}}{d^3k}&=& -2\frac{\alpha_s\,C_F}{\pi^2}\text{Re}\Biggl(i\int_0^{L^+}dy^+\int \frac{d^2\k'}{(2\pi)^2} \frac{\k'\cdot\left(\k'-\delta\k\cos\left(\Omega y^+\right)\right)}{\left(\k'-\delta\k\cos\left(\Omega y^+\right)\right)^2}\nonumber \\ &\times&
\frac{\P(\k'-\bbkappa,L^+-y^+)}{2k^+}
\exp\Biggl\{{(1-i)\frac{\k'^2}{k_f^2}\tan\left(\Omega y^+\right)}\Biggr\}\nonumber
\\
&\times&
\Biggl[1-\exp\Biggr\{i\frac{\left(\k'-\delta\k\cos(\Omega y^+)\right)^2}{2k^+\Omega\sin(\Omega y^+)\cos(\Omega y^+)}
\Biggr\}
\Biggr]\Biggr)\,.
\eeq

\item [{\bf (ii)}] $\textbf{``\emph{Bef-out}"}$: this term can be written as follows
\beq
\label{eq:befout-exact}
\hspace{-2cm}k^+ \frac{dN_{bef-out}}{d^3k}
&=&-2\frac{\alpha_s C_F}{\pi^2}
\text{Re}\Biggl\{
\frac{\bbkappa\cdot\bigl(\bbkappa-\delta\k \cos (\Omega L^+)\bigr)}{\bbkappa^2 \bigl(\bbkappa-\delta\k \cos (\Omega L^+)\bigr)^2}
\exp\Biggl[(1-i)\frac{\bbkappa^2}{2k_f^2}\tan(\Omega L^+)\Biggr]\nonumber\\
&\times&\Biggl(1-\exp\Biggl[i\frac{\bigl(\bbkappa-\delta\k\cos (\Omega L^+)\bigr)^2}{2k^+\Omega\,\sin (\Omega L^+)\cos (\Omega L^+)}
\Biggr]
\Biggr)
\Biggr\}\ . 
\eeq
\end{itemize}
The asymptotic behaviour of both terms \eqref{eq:befin-comp} and \eqref{eq:befout-exact} is determined by how large or small is the argument of the phases. As a consequence two different limits arise: the coherent and the incoherent regime~\cite{Armesto:2013fca}. Despite their differences, both regimes share similar features. Notice that the vacuum coherence pattern is recovered when taking the limit $\hat q \to 0$ in Eqs. (4)-(9)~\cite{Armesto:2013fca}.

\subsection{Coherent regime: $L^+\ll\tau_f$}

This limit corresponds to the physical situation when the emitted gluon remains coherent during all the time while crossing the QCD medium~\cite{Armesto:2013fca}. The medium acts as a unique scattering center and thus the effective momentum transfer is $Q_s^2=\hat q L^+$. The gluon spectrum is reduced to the ``\emph{bef-bef}", ``\emph{bef-out}" and ``\emph{out-out}" terms (see Fig.\eqref{fig:spec}). Interferences are completely suppressed when $|\delta\k|\ll Q_s$ so there are gluon emissions outside of the angle associated to the hard scattering $\theta_{qq}$ and hence, antiangular ordering. The typical momentum scale transferred by the medium $Q_s^2=\hat q L^+$ sets up a upper bound for large angle radiation, so angular emissions lie in the range $\theta_{qq}\leq\theta\leq\theta_{max}$ where $\theta_{max}\sim Q_s/k^+$. The spectrum is completely suppressed above $\theta_{max}$ and the vacuum coherence pattern is recovered, i.e., when $|\delta\k|\gg Q_s$ gluons are radiated in a coherent manner as in the vacuum~\cite{Armesto:2013fca}. 

\subsection{Incoherent regime: $L^+\gg\tau_f$}

In this limit, all the components of the spectrum (see Fig.\eqref{fig:spec}) must be considered. In a similar manner to the coherent regime, there is a partial suppression of interferences between both emitters due to the multiple scatterings in the medium.  Large angle gluon emissions are expected in the kinematic region $\theta_{qq}\leq\theta\leq \theta_{max}$ and the vacuum coherence pattern is again reestablished  for $|\k|>Q_s$.

\section{Single gluon production in pA collisions beyond the eikonal approximation}
\label{sec:pA}

In the studies of particle production at high energies it is based on the eikonal approximation. Phenomenological studies of such calculations have been focused on particle production of large but finite energies. Therefore, an improvement on the kinematical description which captures possible deviations from the high energy regime becomes relevant when comparing with experimental data. Some of the deviations of the high energy limit are possible to study by generalizing our setup to the case of pA collisions. In this section we present some of these new developments in the dilute-dense regime. Among our main results, we obtain a factorized formula in coordinate space for the inclusive gluon spectrum which encodes in a compact manner the non eikonal corrections associated to finite size effects of the target as well as changes in the transverse position of the gluon trajectory~\cite{TNGMS}. Compared with the configuration discussed in the previous section, we do not consider here a hard scattering with a virtual photon.  

When considering proton-nucleus collisions with a particular impact parameter $B$, the proton is usually described by a classical color current~\cite{JalilianMarian:2005jf}
\beq
J^{\mu}_a(x)=\delta^{\mu -}\: \delta(x^-)\;\U^{ab}(x^+,-\infty,\x)\;\;  \rho^{b}(\x\!-\!\B),
\eeq
where $\rho^{b}$ is the transverse density of color charges inside the proton before it reaches the nucleus, and $\U^{ab}(x^+,-\infty,\x)$ is the Wilson line implementing the color precession of those color charges in the background field ${\cal A}^{-}_a(x^+,\x)$. The target is described by a background field ${\cal A}^{\mu}_a(x)=\delta^{\mu -}\: {\cal A}^{-}_a(x^+,\x)$ located in a finite region $x^+\in [0, L^+]$. With these assumptions, we calculate the scattering amplitude ${\cal M}^a_\lambda (k^+,\k, \B)$ by following the procedure discussed in Sect. \ref{sec:CYM}. After performing some calculations the single inclusive spectrum for pA collisions with impact parameter $B$ reads as~\cite{TNGMS}
\beq
\label{pAspec}
\hspace{0cm}&&(2\pi)^3\, (2k^+)\, \frac{dN}{dk^+\, d^2\k} (\B)\!\!= \!\frac{1}{\pi^2}\int_{\y,\z,\yb,\zb}\hspace{-0.6cm}\kappa (\yb,\zb,\y,\z)\Bigg\langle\Biggl\{
\int_{\zb'}\hspace{-0.2cm}e^{-i\k\cdot(\zb-\zb')} \G^{\dagger}_{k^+}(L^+,\zb';0,\zb)\nonumber\\
&&\hspace{1cm}-\U^{\dagger}(L^+,0,\yb)-\frac{1}{2ik^+}\int_{\zb',\bar{y}^+}\hspace{-0.6cm}e^{-i\k\cdot(\zb-\zb')} \U^{\dagger}(\bar{y}^+,0,\yb)\partial^2_{\zb}\left[\G_{k^+}^{\dagger}(L^+,\zb';\bar{y}^+,\zb)\right]
\Biggr\}^{ab}\nonumber\\
&&\hspace{1cm}\times\Biggl\{
\int_{\z'}\hspace{-0.2cm}e^{i\k\cdot(\z-\z')} \G_{k^+}(L^+,\z';0,\z)-\U(L^+,0,\y)\nonumber\\
&&\hspace{1.2cm}+\frac{1}{2ik^+}\int_{\z',y^+}e^{i\k\cdot(\z-\z')} \partial^2_{\z}\left[\G_{k^+}(L^+,\z';y^+,\z)\right]
\U(y^+,0,\y)
\Biggr\}^{bc}\Bigg\rangle_A\nonumber\\
&&\hspace{5.5cm}\times \left\langle\rho^a(\yb\!-\!\B)\rho^c(\y\!-\!\B)\right\rangle_{p}
\eeq
where $\kappa (\yb,\zb,\y,\z)$ is the kernel defined as
\beq
\kappa (\yb,\zb,\y,\z)=e^{i\k\cdot(\zb-\z)}\frac{(\zb-\yb)}{(\zb-\yb)^2}\cdot\frac{(\z-\y)}{(\z-\y)^2}\,,
\eeq
and $\G$ is the gluon propagator in the background field which is given by the following path integral 
\beq
\mathcal{G}^{ab}_{k^+}\left(x^+,\x; y^+,\y\right) = \int_{\r(y^+)=\y}^{\r(x^+)=\x}\mathcal{D} {\bs r} \, \exp\left[i\frac{k^+}{2}\int_{y^+}^{x^+} \!\!d\xi \, \dot{{\bs r}}^2(\xi) \right] \U_{ab}(x^+,y^+) \,.
\eeq
The non-eikonal corrections to the single gluon spectrum \eqref{pAspec} are encoded in the gluon propagator $\G$. Eq. \eqref{pAspec} is written in a factorized form, i.e., it is possible to separate the gluon spectrum into a convolution of three components: the probability to emit a soft gluon given by the kernel $\kappa (\yb,\zb,\y,\z)$, the probability that the emitted gluon scatters with the partons of the target and the color correlator of the projectile charge density~\cite{TNGMS}. As a consistency check,  Eq. \eqref{pAspec} reproduces the well known expression of the single gluon spectrum in the high energy limit. In addition, it is possible to quantify precisely the non eikonal deviations to the inclusive gluon spectrum by performing an eikonal expansion of the gluon propagator in a background field. A more complete description of these new results can be found in Ref.~\cite{TNGMS}. In addition, in Ref.~\cite{TNGMS} we also discuss the effect of non-eikonal corrections in certain spin asymmetries for pA collisions. 

\section{Conclusions}
\label{Concl}

In this contribution we review our recent work on the modifications to the interference pattern between the initial and final state radiation due to the presence of a QCD medium~\cite{Armesto:2013fca}. We obtain an analytical expresion for the medium induced gluon spectrum by using semi-classical methods of pQCD. The gluon spectrum is composed by the independent gluon emissions associated with the incoming and outgoing parton as well as the interference terms between both emitters. The color connection between the initial and final state depends on the time-dependent transverse distance of the quark-gluon system. To shed light on the structure of the spectrum, we consider two asymptotic limits of the gluon spectrum: the incoherent and coherent regime. In both limits we observe a partial loss of coherence due to the medium interactions. Therefore the presence of a QCD medium opens the phase space for large angle emissions, e.g. antiangular ordering, up to a maximal angle $\theta_{max}=Q_s/k^+$ determined by the medium properties. The vacuum coherence is recovered for angle emissions above $\theta > \theta_{max}$.

We present some new preliminary results of the generalization of our setup to pA collisions by including finite width effects as well as non-eikonal corrections to the gluon trajectory inside the target. We are able to write the the single inclusive spectrum in a factorized form in coordinate space. We recover the single inclusive gluon spectrum in the eikonal approximation. We comment that within this approach, it is also possible to study the impact of non-eikonal corrections to the single inclusive cross section in pA collisions~\cite{TNGMS}. 

Our results may have relevant phenomenological consequences on hadron production in nuclear collisions. For instance, for particle production associated to jets in the forward and backward regions where the interferences between initial and final state radiation are absent. At theoretical level, it is not clear which kind of factorisation may be valid in the regime where usual collinear factorisation plausibly fails, and if there are connections with other formalisms such as the hybrid formalism. Future developments are likely to bring new and exciting results. 

\section*{Acknowledgments}

The author thanks to T. Altinoluk, N. Armesto, G. Beuf, H. Ma, Y. Mehtar-Tani and C. Salgado for the fruitful collaboration and for valuable discussions. This work was supported by European Research Council grant HotLHC ERC-2011-StG-279579; by Ministerio de Ciencia e Innovaci\'on of Spain under projects FPA2008-01177, FPA2009-06867-E and FPA2011-22776; by Xunta de Galicia (Conseller\'{\i}a de Educaci\'on and Conseller\'\i a de Innovaci\'on e Industria - Programa Incite); by the Spanish Consolider-Ingenio 2010 Programme CPAN and by FEDER and by U.S. Department of Energy under Grant No.~\rm{DE-SC0004286}.

\section*{References}
\bibliography{IS2013}

\end{document}